\input phyzzx.tex
\tolerance=1000
\voffset=-0.0cm
\hoffset=0.7cm
\sequentialequations
\def\rl{\rightline}

\def\t1{{\tilde 1}}

\def\t{\theta}

\REF{\STR}{J. M. Maldacena, [arXiv:hep-th/9607235],
A. W. Peet, [arXiv:hep-th/0008241]; J. R. David, G. Mandal and S. R. Wadia, Phys. Rep. {\bf 369} (2002) 549, [arXiv:hep-th/0203048].}
\REF{\ATT}{A. Dabholkar, A. Sen and S. P. Trivedi, JHEP {\bf 0701} (2007) 096, [arXiv:hep-th/0611143].} 
\REF{\BEK}{J. Bekenstein, Lett. Nuov. Cimento {\bf 4} (1972) 737; Phys Rev. {\bf D7} (1973) 2333; Phys. Rev. {\bf D9} (1974) 3292.}
\REF{\HAW}{S. Hawking, Nature {\bf 248} (1974) 30; Comm. Math. Phys. {\bf 43} (1975) 199.}
\REF{\CAR}{J. L. Cardy, Nucl. Phys. {\bf B463} (1986) 435.}
\REF{\LEN}{L. Susskind, [arXiv:hep-th/9309145].}
\REF{\SBH}{E. Halyo, A. Rajaraman and L. Susskind, Phys. Lett. {\bf B392} (1997) 319, [arXiv:hep-th/9605112].}
\REF{\HRS}{E. Halyo, B. Kol, A. Rajaraman and L. Susskind, Phys. Lett. {\bf B401} (1997) 15, [arXiv:hep-th/9609075].}
\REF{\EDI}{E. Halyo, Int. Journ. Mod. Phys. {\bf A14} (1999) 3831, [arXiv:hep-th/9610068]; Mod. Phys. Lett. {\bf A13} (1998), [arXiv:hep-th/9611175].}
\REF{\DES}{E. Halyo, [arXiv:hep-th/0107169].}
\REF{\UNI}{E. Halyo, JHEP {\bf 0112} (2001) 005, [arXiv:hep-th/0108167]; [arXiv:hep-th/0308166].}
\REF{\EDIW}{E. Halyo. [arXiv:1403.2333].}
\REF{\WAL}{R. M. Wald, Phys. Rev. {\bf D48} (1993) 3427, [arXiv:gr-gc/9307038]; V. Iyer and R. M. Wald, Phys. Rev. {\bf D50} (1994) 846, [arXiv:gr-qc/9403028]; Phys. Rev. {\bf D52} (1995) 4430, [arXiv:gr-qc/9503052].}
\REF{\EDIH}{E. Halyo, [arXiv:1406.5763].}
\REF{\CARL}{S. Carlip, Phys. Rev. Lett. {\bf 82} (1999) 2828, [arXiv:hep-th.9812013]; Class. Quant. Grav. {\bf 16} (1999) 3327,
[arXiv:gr-qc/9906126].}
\REF{\SOL}{S. Solodukhin, Phys. Lett. {\bf B454} (1999) 213, [arXiv:hep-th/9812056].}
\REF{\DL}{G. A. S. Dias and J. P. S. Lemos, Phys. Rev. {\bf D74} (2006) 044024, [arXiv:hep-th/0602144].}
\REF{\DGM}{S. Das, A. Ghosh and P. Mitra, Phys. Rev. {\bf D63} (2001) 024023,[arXiv:hep-th/0005108].}
\REF{\HSS}{M. Hotta, K. Sasaki and T. Sasaki, Class. Quant. Grav. {\bf 18} (2001) 1823, [arXiv:gr-qc/0011043].}
\REF{\MIP}{M. I. Park, Nucl. Phys. {\bf B634} (2002) 339, [arXiv:hep-th/0111224].}
\REF{\CARLI}{S. Carlip, Phys. Rev. Lett. {\bf B88} (2002) 241301, [arXiv:gr-qc//0203001].}
\REF{\GP}{A. Giacomini and N. Pinamonti, JHEP {\bf 0302} (2003) 014, [arXiv:gr-qc/0301038].}
\REF{\DGW}{O. Dreyer, A. Ghosh and J. Wisniewski, Class. Quant. Grav. {\bf 18} (2001) 1929, [arXiv:hep-th/0101117].}
\REF{\CPP}{M. Cvitan, S. Pallua and P. Prester, Phys. Rev. {\bf D70} (2004) 084043, [arXiv:hep-th/0406186].} 
\REF{\KKP}{G. Kang, J. I. Koga and M. I. Park, Phys. Rev. {\bf D70} (2004) 024005, [arXiv:hep-th/0402113].}
\REF{\CHU}{H. Chung, Phys. Rev. {\bf D83} (2011) 084017, [arXiv:1011.0623].} 
\REF{\SIL}{S. Silva, Class. Quant. Grav. {\bf 19} (2002) 3947, [arXiv:hep-th/0204179].}
\REF{\MAJ}{B. R. Majhi and T. Padmanabhan, Phys. Rev. {\bf D85} (2012) [arXiv:1111.1809]; Phys. Rev. {\bf D86} (2012) 101501,
[arXiv:1204.1422].}
\REF{\KER}{M. Guica, T. Hartman, W. Song and A. Strominger, Phys. Rev. {\bf D80} (2009) 124008, [arXiv:0809.4266]; I. Bredberg,
C. Keeler, V. Lysov and A. Strominger, Nucl. Phys. Proc. Supp. {\bf 216} (2011) 194, [arXiv:1103.2355].}
\REF{\TRA}{D. Birmingham, K. S. Gupta and S. Sen, Phys.Lett. {\bf B505} (2001) 191, [arXiv:hep-th/0102051]; 
A. J. M. Medved, D. Martin and M. Visser, Phys.Rev. {\bf D70} (2004) 024009 [arXiv:gr-qc/0403026].} 
\REF{\BTZ}{A. Strominger, JHEP {\bf 9802} (1998) 009, [arXiv:hep-th/9712251].}
\REF{\FAT}{J. Maldacena and L. Susskind, Nucl. Phys. {\bf B475} (1996) 679, [arXiv:hep-th/9604042].}
\REF{\DIL}{M. Cadoni and P. Carta, Phys. Lett. {\bf B522} (2001)126, [arXiv:hep-th/0107234]; D. Grumiller and R. McNees, JHEP {\bf 04} (2007) 074, [arXiv:hep-th/0703230].}
\REF{\ASH}{I. Mandal and A. Sen, Class. Quant. Grav. {\bf 27} (2010) 214003, [arXiv:1008.3801]; A. Dabholkar, Lect. Notes Phys.
{\bf 851} (2012) 165, [arXiv:1208.4814].}
\REF{\SON}{E. Halyo, in preparation.}
\REF{\RN}{T. Hartman, K. Murata, T. Nishioka and A. Strominger, JHEP {\bf 0904} (2009) 019, [arXiv:0811.4393].}
\REF{\NEX}{A. Castro and F. Larsen, JHEP {\bf 0902} (2009) 037, [arXiv:0908.1121]; A. Castro, A. Maloney and A. Strominger, Phys.Rev. {\bf D82} (2010) 024008, [arXiv:1004.0996].}
\REF{\SONN}{E. Halyo, work in progress.}
\REF{\RIN}{D. A Lowe and A. Strominger, Phys. Rev. {\bf D51} (1995) 1793, [arXiv:hep-th/9412215].}
\REF{\SUN}{A. Giveon and N. Itzhaki, JHEP {\bf 1309} (2013) 079, [arXiv:1305.4799]; JHEP {\bf 1212} (2012) 094, [arXiv:1208.3930].}
\REF{\CIG}{T. Mertens, H. Verschelde and V. I. Zakharov, Phys.Rev. {\bf D82} (2010) 024008, [arXiv:1307.3491]; [arXiv:1410.8009].}


\singlespace
\rl{SU-ITP-15/01}
\pagenumber=0
\normalspace
\medskip
\bigskip
\titlestyle{\bf{Black Holes as Conformal Field Theories on Horizons}}
\smallskip
\author{ Edi Halyo{\footnote*{e--mail address: halyo@stanford.edu}}}
\smallskip
\centerline {Department of Physics} 
\centerline{Stanford University} 
\centerline {Stanford, CA 94305}
\smallskip
\vskip 2 cm
\titlestyle{\bf ABSTRACT}
We show that any nonextreme black hole can be described by a state with $L_0=E_R$ in a $D=2$ chiral conformal field theory with central charge $c=12E_R$ where $E_R$ is the dimensionless Rindler energy of the black hole. The theory lives in the very near horizon region, i.e. around the origin of Rindler space. Black hole hair is the momentum along the Euclidean dimensionless Rindler time direction. As evidence, we show that $D$--dimensional Schwarzschild black holes and $D=2$ dilatonic ones that are obtained from them by spherical reduction are described by the same conformal field theory states.

\singlespace
\vskip 0.5cm
\endpage
\normalspace

\centerline{\bf 1. Introduction}
\medskip

String theory provides a precise counting of the microscopic states that give rise to the entropy of BPS and nonsupersymmetric extreme black holes[\STR]. This is due to either supersymmetry or the attractor mechanism that protects the number of states under extrapolations from weak to strong coupling[\ATT]. Unfortunately, there is no such mechanism for black holes far from extremality. Ideally, we would
like to count nonextreme black hole entropy by identifying the fundamental, microscopic degrees of freedom.
Since this seems to be too hard, we may try to find an effective but microscopic theory that reproduces the Bekenstein--Hawking entropy[\BEK,\HAW]
$$S_{BH}={A_h \over{4 G}} \quad. \eqno(1)$$
Even without knowing the full details of the microscopic theory, this may lead to a
deeper understanding of the fundamental structure black holes. For example, if the horizon degrees of freedom were described by a conformal field theory (CFT) the entropy of a black hole state would be given by the Cardy formula[\CAR] 
$$S_{CFT}=2 \pi \sqrt{{{cL_0} \over 6}} \quad, \eqno(2)$$
where $c$ is the central charge of the CFT and $L_0$ is the conformal weight of the state. Computing $S_{CFT}$ does not require
any knowledge about the CFT beyond $c$ and $L_0$. Of course, if we do not know the details of the CFT, we cannot describe the microscopic degrees of freedom on the horizon. Nevertheless, if the horizon corresponds to a CFT state, it is described by a field theory (and not gravity) that is severely constrained by $c$, $L_0$ and conformal symmetry.

It has been known for some time that black hole entropy is given by the simple relation $S_{BH}=2 \pi E_R$ where $E_R$ is the dimensionless Rindler energy obtained from the near horizon geometry of the black hole[\LEN]. This formula has been
used to compute the entropy of a wide variety of black holes and de Sitter space[\SBH-\UNI]. In ref. [\EDIW], it was shown that the same relation holds beyond General Relativity, in generalized theories of gravity. $E_R$ is exactly Wald's Noether charge $Q$ and therefore Wald entropy[\WAL] can be written as
$$S_{Wald}=2 \pi Q=2 \pi E_R \quad. \eqno(3)$$
Even though the relationship of eq. (3) to holography is not manifest, it can be shown that $E_R$ is a holographic quantity that is given by the surface Hamiltonian of the gravitational theory evaluated on the horizon[\EDIH]. 
The simplicity and generality of eq. (3) certainly requires a microscopic explanation. On the other hand, the striking similarity between eqs. (2) and (3) strongly hints that Wald entropy might arise from a CFT description of nonextreme black holes.

In this paper, we show that nonextreme black holes can be described by $D=2$ chiral CFTs that live on their horizons, i.e. on Rindler space. These CFTs have a central charge of $c=12E_R$ and the black hole corresponds to a state with $L_0=E_R$.
Due to the exponential map from Euclidean to (Euclidean) Rindler space, there is a shift of $-c/24=-E_R/2$ in the eigenvalues of $L_0$. Using these results, the Cardy formula gives the correct Wald entropy as $S_{Wald}= 2 \pi E_R$. 
In addition, we show that $D$--dimensional Schwarzschild black holes and $D=2$ dilatonic black holes that effectively describe them
have precisely the same CFT description. Since $D=2$ gravity
theories are described by CFTs, this provides additional support for our picture.
We identify the black hole hair as units of momentum along the dimensionless Euclidean Rindler time circle. Unfortunately, since we do not know the details of these CFTs, we are not able to describe the degrees of freedom that carry this hair.

Starting with ref. [\CARL], similar descriptions of black holes have appeared in the literature[\CARL-\MAJ]. In this paper, our approach is completely new and our description, hopefully, is more intuitive than those in previous attempts. In [\CARL-\MAJ],
one looks for diffeomorphisms that keep the horizon invariant and finds the charges associated with them. These charges are either
Wald's Noether charge, $Q$, or the surface Hamiltonian, $H_{sur}$, that gives rise to surface deformations of the horizon. One then quantizes the charges and obtains a Virasoro algebra with a central charge. For the black hole state of the CFT, $c/12=L_0$ where both quantities are given by either $Q$ or $H_{sur}$. The connection with our result becomes apparent when one realizes that both $Q$ and $H_{sur}$ are precisely $E_R$[\EDIW,\EDIH]. We refer the reader to refs. [\CARL-\MAJ] for the details.

Our description of nonextreme black holes by horizon CFTs seems very similar to the Kerr/CFT correspondence[\KER] which describes
extreme Kerr black holes by chiral CFTs with $c/12=L_0=J$ where $J$ is the angular momentum of the black hole. This is exactly our picture with $E_R$ replaced by $J$. Of course, in our case the black holes are nonextreme and the relationship between the two 
horizon CFTs is not trivial.

This paper is organized as follows. In the next section we review the near horizon description of black holes and the dimensionless
Rindler energy, $E_R$. In section 3, we show that nonextreme black holes can be described by states of $D=2$ chiral CFTs that live on their horizons. We determine the central charge of the CFT, the conformal weight of the black hole state and the shift in $L_0$.
In section 4, we support our picture by showing that $D$--dimensional Schwarzschild black holes and $D=2$ dilatonic black holes
that effectively describe them have exactly the same description in terms of CFT states. Section 4 contains a discussion of our results and our conclusions.

\bigskip
\centerline{\bf 2. Near Horizon Description of Black Holes and Rindler Energy}
\medskip

In this section, we review the well-known fact that the near horizon geometry of any nonextreme black hole is Rindler space and
the derivation of the dimensionless Rindler energy, $E_R$, which gives the Wald entropy. 
Since the near horizon metrics of all nonextreme black holes are the same (up to rescalings of time), 
the only property that distinguishes between different black holes is $E_R$ that is conjugate to Rindler time.
As we show in the next section, $E_R$ plays a crucial role in the description of black holes in terms of CFTs on their horizons. 
The central charges of the CFTs and the conformal weights of the black hole states are both determined by $E_R$.

Consider any nonextreme black hole with a generic metric of the form
$$ ds^2=-f(r)~ dt^2+ f(r)^{-1} dr^2+ r^2 d \Omega^2_{D-2} \quad, \eqno(4)$$
in D-dimensions. The horizon is at $r_h$ which is determined by
$f(r_h)=0$. If in addition, $f^{\prime}(r_h) \not =0$, the near horizon geometry is described by Rindler space. Near the
horizon, $r=r_h +y$ with $y<<r_h$,
which leads to the near horizon metric
$$ds^2=-f^{\prime}(r_h)y~ dt^2+(f^{\prime}(r_h)y)^{-1} dy^2+ r_h^2 d \Omega^2_{D-2} \quad. \eqno(5)$$
In terms of the proper radial distance, $\rho$, obtained from $d\rho=dy/\sqrt{f^{\prime}(r_h)y}$ the metric becomes
$$ds^2=-{f^{\prime 2}(r_h) \over 4} \rho^2 dt^2+d \rho^2+ r_h^2 d \Omega^2_{D-2} \quad. \eqno(6)$$
Using the dimensionless Euclidean Rindler time $\tau=i(f^{\prime}(r_h)/2)~ t$  we find
$$ds^2=\rho^2 d \tau^2 + d \rho^2 + r_h^2 d \Omega^2_{D-2} \quad, \eqno(7)$$
where the metric in the $\tau$--$\rho$ directions describes Rindler space.
We stress that the Euclidean Rindler space in eq. (7), which naively looks like flat space in polar coordinates, is not the same as the flat Euclidean space. These two spaces are
related by an exponential coordinate transformation which plays an important role in the horizon CFT description of black holes as we show in the next section. 

In Euclidean signature, Rindler coordinates that describe Rindler observers
at fixed $\rho$ are related to the flat coordinates, $X$ and $T$, that describe freely falling observers by
$$T=\rho~ sin~ \tau \qquad X=\rho~ cos~ \tau \quad. \eqno(8)$$
(where the usual factors of the surface gravity are missing since $\kappa=1$ when we use the dimensionless Rindler time $\tau$).
We see that $\tau$ translations are rotations in the $T$--$X$ plane and the $\tau$ direction has a period of $\beta=2 \pi$. 
The horizon which is at the origin, $\rho=0$, of Rindler space.
World--lines of Rindler observers at constant $\rho$ are circles of radius $\rho$. The proper energy and temperature measured by these observers are $E_R/\rho$ and $1/2 \pi \rho$ respectively.

In Ref.[\LEN] the dimensionless Rindler energy $E_R$ conjugate to $\tau$ was calculated by using the Poisson bracket
{\footnote1{The i on the left--hand side is due to the Euclidean signature for time.}}
$$i=\{E_R,\tau\}=\left({\partial E_R \over \partial M}{\partial \tau \over \partial t}-{\partial E_R \over \partial t}
{\partial \tau \over \partial M} \right) \quad, \eqno(9)$$
where $M$ is the mass of the black object conjugate to $t$. Taking $E_R$ to be time independent which is a good approximation for large enough black holes we get
$$dE_R={2 \over f^{\prime}(r_h)}~ dM \quad.\eqno(10)$$
Using the definition of Hawking temperature obtained from the metric, $T_H=f^{\prime}(r_h)/4 \pi$, eq. (10) can be written as
$$d(2 \pi E_R)={dM \over T_H} \quad. \eqno(11)$$
We conclude that $S=2 \pi E_R$ for all nonextreme black holes since eq. (11) is the First Law of Thermodynamics. 
The arguments above are completely general and apply beyond General Relativity. In fact it can be shown that $E_R$ is exactly 
Wald's Noether charge $Q$ and therefore gives Wald entropy as in eq. (3)[\EDIW].

The metric in eq. (7) is the same for all nonextreme black holes in any theory of gravity. All the information about a particular black hole is now contained in the relation between the Schwarzschild time $t$ and the dimensionless Euclidean Rindler time $\tau$,
i.e. in $\tau=i(f^{\prime}(r_h)/2)t=i\kappa t$ where $\kappa$ is the surface gravity. Alternatively, and more relevant for
our purposes, the information about the black hole resides in the dimensionless Rindler energy $E_R$ which is canonically conjugate to $\tau$ by eq. (9). $E_R$ is not only the dimensionless energy associated with the black hole.
If we take $\tau$ to be an angle in eq. (7), then
$E_R$ can also be interpreted as the angular momentum of the black hole state (not to be confused with the angular momentum 
of the black hole in real space). 

To summarize, a nonextreme black hole with mass $M$ is described, in the near horizon region, by the Euclidean Rindler space with
metric in eq. (7) and the dimensionless Rindler energy, $E_R(M)$. The horizon is at $\rho=0$ and the dimensionless Rindler
temperature of the black hole, i.e. the inverse periodicity of the compact $\tau$ direction is $1/2 \pi$. $E_R$ also represents the 
angular momentum of the black hole in Rindler space.

\bigskip
\centerline{\bf 3. Black Holes as CFTs on Horizons}
\medskip

The simplicity and generality of the relation between the Wald entropy and $E_R$ given by eq. (3) clearly requires an explanation.
$E_R$ is a classical quantity obtained from the metric and therefore does not tell us what the microscopic black hole
degrees of freedom are. In ref. [\LEN], $E_R$ was identified with the square root of the string oscillator number, 
$E_R=\sqrt{n}$, so that the string entropy (assuming the central charge $c=6$ in all cases)
$S_{string} =2 \pi \sqrt{n}$ gives the black hole entropy. In this picture, a black hole
is described by a very massive, wildly oscillating and very long string at its Hagedorn temperature. The mass of the string is much larger than that of the 
black hole, $M$, i.e. $m_{string} \sim \sqrt{n} / \ell_s \sim GM^2/\ell_s >> M$. 
{\footnote2{For simplicity, all formulas in this paragraph refer to the $D=4$ Schwarzschild black hole.}}
An asymptotic observer
measures the correct black hole mass due to the gravitational redshift of mass between the near horizon region and asymptotic infinity by the factor $\ell_s/GM$. The Hawking temperature is the redshifted Hagedorn temperature $T_H \sim (\ell_s/GM)(1/\ell_s) \sim 1/GM$.
A string with such a large $n$ is very long, with length $\sim \sqrt{n} \ell_s$ and covers the black hole horizon due to its transverse oscillations. One then gets one string bit per Planck area on the horizon[\UNI].
The string tension is redshifted to a very small value, $\sim 1/G^2M^2 << 1/\ell_s^2$ so the string does not look like a fundamental one to an asymptotic observer. 

In this paper, we propose a different microscopic interpretation of $E_R$ and eq. (3). 
We show that (the near horizon region of) a nonextreme black hole can be described by a $D=2$ chiral CFT
living on Rindler space with $c=12E_R$. The black hole corresponds to a CFT state with 
$L_0=E_R$. The black hole hair is given by the momentum along the  dimensionless Euclidean Rindler time.

Consider the near 
horizon region $\rho<<1$ in the Rindler metric of eq. (7) what we will call the ``very near horizon" region. 
The local energy $E_R/\rho$ becomes arbitrarily large as $\rho \to 0$. In this limit, all dimensionful parameters can be ignored and
the near horizon physics is described by a CFT[\CARL,\SOL,\HSS,\CHU].
It can also be shown that in the same limit, the transverse directions decouple from Rindler space[\TRA].
Intuitively, this is most easily seen from the metric in eq. (4) where as $r \to r_h$, $f(r) \to 0$ and therefore the radial direction gets arbitrarily long whereas the transverse ones are fixed. Thus, it is much easier to excite modes along the radial direction and they dominate the near horizon physics.
To summarize, a $D=2$ CFT on Rindler space is the most general description of the near horizon region of a black hole.

More specifically, we would like to describe the near horizon region of a black hole with the Euclidean Rindler metric in eq. (7), dimensionless Rindler temperature $T_R=1/2 \pi$ and dimensionless Rindler energy $E_R$ in terms of a $D=2$ CFT. 
We first argue that the CFT has to be chiral, i.e. with only left--movers. The Euclidean Rindler space has a $U(1)$ isometry
which is the invariance along the Euclidean Rindler time circle. In all description of black holes in terms of CFTs, it is either 
a $U(1)$[\KER] or an $SL(2,R)$[\BTZ] isometry of the near horizon region that gets enhanced to an asymptotic Virasoro algebra. Since Rindler space has only one $U(1)$ we expect one Virasoro algebra giving rise to a chiral CFT. We note that, in the Rindler case, the Virasoro algebra is not asymptotic but realized in the ``very near horizon" region[\CARL-\MAJ].
In addition, we found above that, in Rindler space, $E_R$ is both
the dimensionless energy and angular momentum of the black hole state. Thus, at least classically,
$$H=L_0+{\bar L}_0=J=L_0-{\bar L}_0=E_R \quad. \eqno(12)$$ 
This describes a chiral CFT with ${\bar L}_0=0$ and $L_0=E_R$.
As a result, we identify the eigenvalue of $L_0$ with $E_R$. {\footnote3{The eigenvalue of $L_0$ is the scale dimension of operator and state in the CFT that correspond to the black hole.}} Finally, we can also motivate the
chirality of the CFT by noting that time evolution in the metric in eq. (7) corresponds to rotations and therefore 
right--movers would go backwards in time.

A state in a chiral CFT with central charge $c$ is defined by its conformal weight for left--moving modes $L_0$. Depending on the space the CFT lives on, the eigenvalues of $L_0$ may shift by a constant (relative to those on the Euclidean plane).
In fact, in our case, there must be a shift if we take the Euclidean vacuum to be at zero energy. 
Then, the conformal weights on Rindler space are shifted due to the exponential map between the Euclidean and (Euclidean) Rindler spaces. Defining $\rho=\kappa^{-1}exp(\kappa \xi)$ in eq. (8)
{\footnote4{Here, for clarity, we restore $\kappa$ in the transformations. In our case, $\kappa=1$ and it will be set to unity in eq. (16) below.}} we can write the map from Euclidean to Rindler space as
$$ z=\kappa^{-1} exp(\kappa {u}) \quad, \eqno(13)$$
where $z=X +iT$ and ${u}=\xi+i \tau$. Under a coordinate transformation $z=f({u})$ the energy--momentum tensor
transforms as
$$T^{\prime}({u})d{u}^2=T(z)dz^2+{c \over 12}\{{u},z\}du^2 \quad, \eqno(14)$$
where 
$$\{{u},z\}={f^{\prime \prime \prime} \over f^{\prime}}-{3 \over 2} \left(f^{\prime \prime} \over f^{\prime} \right)^2 \eqno(15)$$
is the Schwartzian derivative of the mapping in eq. (13). The Rindler space energy--momentum tensor with the shift is
$$T^{\prime}({u})=T(z)z^2-{c \over 24} \quad. \eqno(16)$$
As a result, we find that in Rindler space conformal weights are given by $L_0^{\prime}=L_0-c/24$.

The Cardy formula for the entropy of a state with weight $L_0^{\prime}$ in a chiral CFT with central charge $c$ is 
$$S=2 \pi \sqrt{{{c L_0^{\prime}} \over 6}} \quad. \eqno(17)$$
Eq. (17) can also be suggestively written as
$$S={\pi^2 \over 3} c T_{CFT} \quad, \eqno(18)$$
where the dimensionless temperature of the CFT is given by
$$T_{CFT}={1 \over \pi} \sqrt{{{6 L_0^{\prime}} \over c}} \quad. \eqno(19)$$
We now identify the Rindler and CFT temperatures, i.e. $T_R=T_{CFT}=1/2 \pi$, and obtain 
$c/12=L_0=E_R$. Substituting these into the Cardy formula we get $S_{CFT}=2 \pi E_R=S_{Wald}$ as required.

We found that the near horizon region of a black hole which is Rindler space is described a $D=2$ chiral CFT on (the same) Rindler space with $c/12=L_0=E_R$ and a shift of $-c/24=-E_R/2$ in the eigenvalue of $L_0$. Using eq. (17), this CFT state gives the correct Wald entropy. 

Unfortunately, we do not know the details of this CFT state beyond its central charge and scaling dimension. In fact, we do not even know if it is unitary or modular invariant even though we used
Cardy's formula that assumes these properties. Turning the argument around, we can claim that this CFT has to be unitary and modular
invariant since Cardy's formula gives the correct black hole entropy. We also note that, strictly speaking, Cardy's formula is only valid asymptotically, i.e. for $L_0>>c$ whereas in our case $c=12 L_0$. It is well--known that this problem is solved by resorting
to fractionation[\FAT] which gives rise to twisted sectors of the CFT. The most highly twisted sector, with a twist $E_R$, dominates the entropy and effectively rescales the central charge and the conformal weight to to $c=12$ and $L_0^{\prime}=E_R^2/2$ 
respectively. These values of $c$ and $L_0$ again give the correct black hole entropy through the Cardy formula.

From the description of the CFT above, we see that the black hole hair is basically momentum along the dimensionless Euclidean Rindler 
time direction. The total momentum along this direction is $E_R$ (minus the shift by $E_R/2$) and different
ways of building it in the CFT gives rise to the entropy. From the definition $L_0=-z(\partial/\partial z)$ we find that the black hole state in the CFT is given by $\phi = a/z^{E_R}$ where $a$ is a constant. This is simply a state with scaling dimension and 
spin $E_R$ which satisfies
$$(L_0-{E_R \over 2}) \vert \phi>={E_R \over 2} \vert \phi> \quad, \eqno(20)$$
as required. There are many ways to build this state which can be generically written as
$$\vert \phi>=\phi_1 \phi_2 \ldots \phi_{N} \vert0> \quad, \eqno(21)$$
where we assumed that there are a total of $N$ (and not necessarily $12E_R$) CFT operators with a total central charge $12E_R$
acting on the vacuum. Each $\phi_i$ carries a conformal weight of $L_{0i}=n_i$ subject to the constraint $\Sigma_i n_i=E_R$. The conformal weight that each $\phi_i$ carries arises from different combinations of raising operators such that
$$\phi_i=L_{-1}^{n_{i1}}L_{-2}^{n_{i2}} \ldots L_{-n_i} \quad, \eqno(22)$$
subject to the constraint $\Sigma_j n_{ij}=n_i$. Here, $L_{-i}$ are the raising operators and $n_{ij}$ are the number of times they operate. Black hole entropy simply counts the number of these different combinations which can be obtained from the partition of integers.

At this point a number of comments are in order. First, the shift obtained from the mapping in eq. (13) is the only one that reproduces
the correct entropy if we assume that the Rindler and CFT temperatures match. It is gratifying to see that exactly this shift arises from the mapping between Euclidean to (Euclidean) Rindler coordinates. 
Second, the CFT satisfies the First Law of Thermodynamics as it should. Using eq. (11) we find
$$dE=TdS=T \left({{\pi^2 c} \over 3}\right) dT \quad. \eqno(23)$$
Even though $T=1/2 \pi$ is fixed we can vary it by creating deficit angles. Taking the integral of eq. (23) with respect to
the deficit angle and then setting it to zero gives $E=E_R/2$ which precisely the value of $L_0^{\prime}=L_0-c/24$. Alternatively, using eq. (23) to solve for $T(E)$ we get $S=2 \pi \sqrt {c E/6}$ which again shows that $E=L_0^{\prime}$.
Thus, the CFT satisfies the First Law of Thermodynamics by taking the shift in the energy into account.

In some ways it is better to think of the CFT as living on a cylinder rather than on the Euclidean Rindler space. This can be achieved
by the standard exponential mapping from the plane to the cylinder, i.e. $z=exp(2 \pi i v/L)$ where $v$ parametrizes the cylinder and $L$ is the size of the compact direction which corresponds to Euclidean Rindler time. On the cylinder, the shift in the eigenvalue of $L_0$ can be seen as the Casimir effect due to the compact direction. In general, the (dimensionless) energy
of a CFT would shift by $-\pi c/12 L$ where $L$ is the size of the compact direction. In our case, $L$ is the 
periodicity of the dimensionless Rindler time, i.e.  $L=2 \pi$ giving exactly a shift of $-c/24$.
In addition, the horizon which is at the origin of Rindler space breaks conformal invariance. Of all the $L_n$ that
generate the Virasoro algebra only $L_0$ (or rotations and dilatations) leaves the horizon invariant. The exponential
mapping from the Rindler space to the cylinder, maps the origin to $- \infty$ and restores the full conformal symmetry. Finally, the Virasoro generators are given by $L_n=-z^{1+n}(\partial/\partial z$). Taking $z=r e^{i \tau}$ with $r=1$ we find 
$L_n=-e^{i n \tau}(\partial/\partial \tau)$ which satisfy $[L_n,L_m]=i(n-m)L_{n+m}$. This is the diffeomorphism algebra
of the Euclidean Rindler time circle $S^1$, $Diff(S^1)$, which is the source of the (classical) Virasoro algebra, i.e. the CFT.
This $S^1$ is precisely the compact direction of the cylinder.

In the absence of any ideas about the degrees of freedom, we may speculate that
the total central charge of $12 E_R$ suggests that the CFT describes $E_R$ superstrings oscillating in the eight transverse
directions (to the Rindler space). Then $N=16E_R$ in eq. (22) since each type II string comes with eight transverse bosons, $X^i$ and eight transverse fermions $\psi^i$  for a total of $8E_R$ bosons and $8E_R$ fermions. The strings wrap the Euclidean time circle of the cylinder (with length $2 \pi$) and carry momentum along the circle. Since there are no longitudinal modes on the strings, momentum
is carried by the transverse modes $X^i$ and $\psi^i$. If we assume that fractionation[\FAT] takes place at strong coupling, then $E_R$ strings wrapping the circle become one long string of length $2 \pi E_R$. As a result, momentum along the circle is quantized in
units of $1/2\pi E_R$. We end up with one long string with $c=12$ and a rescaled
$L_0^{\prime}=E_R^2/2$. Now, since $L_0^{\prime}>>c$, which is the asymptotic region of parameter space, we can use the 
Cardy formula which gives the right black hole entropy. Even though this picture is similar to the one described at the beginning of section 3 there is an important difference. In the above case the long string is
wound around the Euclidean Rindler time circle. The factor of 2 that was missing in the old picture is made up simply by the
difference between values of $L_0$ and $L_0^{\prime}$.

\bigskip
\centerline{\bf 4. $D=2$ Dilatonic Black Holes and Horizon CFTs}
\medskip

In this section, we would like to support the description of black holes in terms of horizon CFTs by considering
$D$--dimensional Schwarzschild black holes. Using the prescription given in section 3 these black holes can be described by a chiral $D=2$ CFT with $c/12=L_0=E_R$. On the other hand, we can dimensionally reduce gravity on $S^{D-2}$ and obtain $D=2$ dilatonic gravity. This effective theory has black hole solutions that
can also be described by CFTs on their horizons. This is perhaps less surprising since two dimensional
gravity is known to be equivalent to a CFT. In addition, the effects of any matter such as the dilaton is redshifted near the horizon
and will not break conformal symmetry. We show that the horizon CFTs that describe $D=2$ dilatonic black holes are exactly the ones that describe $D$--dimensional Schwarzschild black holes; i.e. $E_R$s in both cases match and therefore the black hole states in both horizon CFTs are the same.

Consider the $D$--dimensional Einstein--Hilbert action
$$A_{EH}={1 \over 16 \pi G_D} \int d^Dx \sqrt{-g_D}R_D \quad, \eqno(24)$$
where $G_D,g_D$ and $R_D$ are the $D$--dimensional Newton's constant, metric and Ricci scalar
respectively. $D$--dimensional Schwarzschild black holes are given by 
$$ds^2 =-\left(1-{\mu \over r^{D-3}}\right)dt^2+\left(1-{\mu \over r^{D-3}}\right)^{-1}dr^2+r^2d\Omega_{D-2} \quad, \eqno(25)$$   
where 
$$\mu={{16 \pi G_D M} \over {(D-2)A_{D-2}}} \qquad A_{D-2}={{2 \pi^{(D-1)/2}} \over {\Gamma((D-1)/2)}} \quad. \eqno(26)$$ 
Following the procedure outlined in section 2, we can find the dimensionless Rindler energy[\SBH]
$$E_R={2 \over {(D-2)}} M^{(D-2)/(D-3)} \left({{16 \pi G_D} \over {(D-2)A_{D-2}}} \right)^{1/(D-3)} \quad, \eqno(27)$$
which gives the correct entropy through $S=2 \pi E_R$. Then, the results of section 3 indicate that $D$--dimensional Schwarzschild black holes are described by $D=2$ chiral CFTs that live on their Rindler horizons with $c/12=L_0=E_R$ where $E_R$ is given by eq. (27).

Now, if we dimensionally reduce the Einstein--Hilbert action over a $(D-2)$--dimensional sphere of radius $\lambda r=\Phi^{-a}$ where
$a=1/(2-D)$, i.e. consider only the s--wave sector of D--dimensional gravity, by using the ansatz
$$ds^2=g_{\mu\nu}dx^{\mu}dx^{\nu}+{\Phi \lambda^{2-D}}d\Omega^2_{D-2} \quad, \eqno(28)$$
where $\mu,\nu=0,1$ and $x^0=t$, $x^1=r$. Note that the radius over which we dimensionally reduce the theory is not fixed but depends on the dilaton, $\Phi$. The constant $\lambda$ is proportional to the inverse Planck length 
$$\lambda=((2(D-2)^{D-2})^{1/(D-3)} \left({{A_{D-2}} \over {16 \pi G_D}} \right)^{1/(D-2)} \quad, \eqno(29)$$
This spherical dimensional reduction gives rise to $D=2$ dilatonic gravity with the action[\DIL]
$$A={1 \over 2} \int d^2x \sqrt{-g}(\Phi R+ \lambda^2 V(\Phi)) \quad. \eqno(30)$$
Here $R$ is the two dimensional Ricci scalar and the dilaton potential is given by $V(\Phi)=(a+1) \Phi^a$. 
The action in eq. (30) has generic black hole solutions with the metric[\DIL]
$$ds^2=-\left(\Phi^{a+1}-{{2M} \over \lambda}\right) dt^2+ \left(\Phi^{a+1}-{{2M} \over \lambda}\right)^{-1} dr^2 \eqno(31)$$
and the linear dilaton $\Phi=\lambda r$. For $a=1/(2-D)$ these correspond to $D$--dimensional Schwarzschild black holes with horizons 
at $\Phi_h^{a+1}=2M/\lambda$. The mass, temperature and entropy of these black holes are given by[\DIL]
$$M={\lambda \over 2} \Phi_h^{a+1} \qquad \qquad T={\lambda \over {4 \pi}}(a+1) \Phi_h^a \qquad \qquad S=2 \pi \Phi_h \quad, \eqno(32)$$
respectively. These precisely match the corresponding quantities for $D$--dimensional Schwarzschild black holes. Thus, we conclude that the essential information about $D$--dimensional black holes is contained in the physics of the $\tau$--$\rho$ directions (or equivalently the s--wave sector) as we claimed in section 3.

It is easy to see that the near horizon geometry of the metric in eq. (31) is two dimensional Rindler space. Therefore, we can
find $E_R$ for this case by applying the procedure outlined in section 2. Near the horizon, $\Phi=\Phi_h+y$ with $y<<\Phi_h$
which leads to the near horizon metric (in Euclidean time)
$$ds^2={{\lambda^2(a+1)^2} \over 4} \Phi_h^{2a} \rho^2 dt^2 + d\rho^2 \quad, \eqno(33)$$
where the proper radial distance is given by
$$\rho= {2 \over {\sqrt{\lambda (a+1)}}} \Phi_h^{a/2} \sqrt{y} \quad. \eqno(34)$$
The temperature obtained from the definition of the dimensionless Rindler time, $\tau=(\lambda (a+1) \Phi_h^a /2) t=2 \pi T t$
agrees with eq. (32). Using eq. (11) we find the Rindler energy $E_R= \Phi_h$. 
Therefore, we can describe the $D=2$ dilatonic black holes by states in a chiral $D=2$ CFT with 
$c/12=L_0=E_R=\Phi_h$. Cardy's formula gives the correct entropy when the $L_0$ shift of $-c/24=-\Phi_h/2$ is taken into account.

Now, the Rindler energy for the $D=2$ dilatonic black holes, $E_R=\Phi_h$ exactly matches that for the $D$--dimensional
Schwarzschild black holes given by eq. (27). Thus the description of both types of black holes in terms of horizon CFTs is
identical. We consider this result as additional evidence for the horizon CFT description of nonextreme black holes.


\bigskip
\centerline{\bf 5. Conclusions and Discussion}
\medskip

We found that any nonextreme black hole can be described by a state in a 
chiral $D=2$ CFT with $c/12=L_0=E_R$ where $E_R$ is the dimensionless Rindler energy. The black hole hair is momentum along the
dimensionless Euclidean Rindler time direction of the near horizon geometry. The entropy arises due to the different number of ways the total momentum can be realized 
in the CFT. Even though we can compute $c$ and $L_0$ we do not know the details of this CFT. Fortunately, this is not necessary to compute the
entropy through Cardy's formula. Needless to say, it is very important to find out exactly what this CFT is. One proposal is that
it is a Liouville theory obtained from compactifying the gravitational theory over the horizon[\SOL,\DL,\GP]. The Liouville
field is basically the radius of the horizon which is taken to be a function of $\tau$ and $\rho$. In this approach, black hole entropy arises from the fluctuations of the horizon radius. In fact, the Liouville theory in question is obtained from
a nontrivial field transformation of the $D=2$ dilatonic gravity in eq. (32).
It would be interesting to investigate the relationship between this Liouville theory and the horizon CFTs.

In this paper we considered nonextreme black holes with near horizon regions that are Rindler spaces. There are non--BPS but extreme black holes, e.g. extreme Reissner--Nordstrom and dyonic black holes which have near horizon geometries of the type $AdS_2 \times S^n$[\ASH]. In fact, in a certain limit, the $AdS_2$ factor of these spaces has a horizon and
can be more accurately described as $AdS_2$ Rindler space. Therefore, we can use the results of this paper to compute the entropy of these extreme black holes[\SON]. We expect that $AdS_2$ Rindler spaces are described by $D=2$ chiral CFTs with $c/12=L_0=E_R$
just like nonextreme black holes. This means that horizon CFTs describe both extreme and nonextreme black holes giving
rise to a uniform description the entropy of all types of black holes[\SON]. In fact, such black holes have been described by dual CFTs in ref. [\RN]. A preliminary comparison shows that the central charge in ref. [\RN] is $Q_e$ times bigger that what our horizon CFT predicts whereas the temperature is $Q_e$ times smaller. It would be interesting to understand the origin of this discrepancy.

As mentioned in the introduction, our results are very similar to those obtained from the extreme Kerr/CFT correspondence[\KER]. In that case, the near horizon region of the
black hole is described by a CFT with $c/12=L_0=J$ where $J$ is the angular momentum of the extreme Kerr black hole. The black hole entropy is then given by the Cardy formula.
This is exactly our result for nonextreme black holes with $E_R$ replaced by $J$. We expect that, if we can describe the near horizon of a generic Kerr black hole by a CFT, its extreme limit will become the Kerr/CFT correspondence whereas the $J=0$ limit will lead to our results. Thus these two limiting descriptions may be continuously connected to each other arising from a more general correspondence.
In ref. [\NEX], the near--extreme Kerr black hole has been described as a CFT that has both left and right moving
sectors. In our case, since the near horizon region is Rindler space the CFT is chiral. However, these two cases are different
since in [\NEX] the deviation from extremality is infinitesimal whereas in our case it is finite[\SONN]. 

If we apply our results to $AdS_{n+2}$ Schwarzschild black holes we find two different descriptions in terms of CFTs. One is the well--known
description of the black hole by a thermal state of the boundary CFT on $R \times S^n$. The other is the $D=2$ horizon CFT that lives
on Euclidean Rindler space that was described in this paper. It would be interesting to see if there is a relationship between these two descriptions.
In general, these two CFTs live on different spaces and they have different central charges, conformal weights, temperatures etc.
even though they both give rise to the same entropy. Perhaps, dimensional reduction on a sphere and the renormalization of the CFT parameters due to the running from the boundary to the horizon may lead to a relationship between these two very different CFTs.

Most importantly, in order to discover the microscopic description of the chiral CFTs discussed above, it is crucial to construct Rindler space in string theory. Any black hole in string theory with large nonextremality is a possible candidate. One approach is to start with the well--understood near--BPS or near--extreme black holes and increase the nonextremality.
Unfortunately, in this regime, it is very hard to follow the states and give a precise microscopic description of the black holes. 
If our speculations at the end of section 3 have any merit, another option would be to study the physics of strings wrapping
Euclidean time circles. Finally, perhaps it is possible to get a microscopic description by examining string theory in Rindler space along the lines of refs. [\RIN,\SUN,\CIG].

\endpage

\bigskip
\centerline{\bf Acknowledgments}

I would like to thank Lenny Susskind for useful discussions and the Stanford Institute for Theoretical Physics for hospitality.

\vfill

\refout

\end
\bye